# Experimental observation of the role of countercations on the electrical conductance of Preyssler-type polyoxometalate nanodevices.

Cécile Huez,[1,‡] Séverine Renaudineau,[2] Florence Volatron,[2] Anna Proust[2,*] and Dominique Vuillaume.[1,*]

1) Institute for Electronics Microelectronics and Nanotechnology (IEMN), CNRS, University of Lille, Av. Poincaré, Villeneuve d'Ascq, France
2) Institut Parisien de Chimie Moléculaire (IPCM), CNRS, Sorbonne Université, 4 Place Jussieu, Paris, France
‡) Now at : Instituto de Ciencia Molecular (ICMoL), University of Valencia, Spain.

* Corresponding authors : anna.proust@sorbonne-universite.fr ; dominique.vuillaume@iemn.fr

**Abstract.**

Polyoxometalates are nanoscale molecular oxides with promising properties that are currently explored for molecule-based memory devices. In this work, we synthesize a series of Preyssler polyoxometalates (POMs), [Na⊂$P_5W_{30}O_{110}$]$^{14-}$, stabilized with four different counterions, $H^+$, $K^+$, $NH_4^+$ and tetrabutylammonium ($TBA^+$), and we study the electron transport properties at the nanoscale (conductive atomic force microscopy, C-AFM) of molecular junctions formed by self-assembled monolayers (SAMs) of POMs electrostatically deposited on ultraflat gold surface prefunctionalized with a positively charged

SAM of amine-terminated alkylthiol chains. We report that the electron transport properties of $P_5W_{30}$-based molecular junctions depend on the nature of the counterions, the low-bias current (in the voltage range [-0.6 V ; 0.6 V]) gradually increasing by a factor ~100 by changing the counterion in the order $K^+$, $NH_4^+$, $H^+$ and $TBA^+$. From a statistical study (hundreds of current-voltage traces) using a simple analytical model for charge transport in nanoscale devices, we show that the energy position of the lowest unoccupied molecular orbital (LUMO) of the $P_5W_{30}$ with respect of the Fermi energy of the electrodes increases from ~ 0.4 eV to ~ 0.7 eV and that that electrode coupling energy also increases from ~ 0.05 to 1 meV in the same order from $K^+$, $NH_4^+$, $H^+$ to $TBA^+$. We discuss several hypotheses on the possible origin of these features, such as a counterion-dependent dipole at the POM/electrode interface and counterion-modulated molecule/electrode hybridization, with, in both cases, the largest effect in the case of $TBA^+$ counterions.

**Keywords.** polyoxometalate, counterion, electron transport, molecular electronics, conductive-AFM.

**Introduction.**

Ions inserted (inadvertently or by a controlled way) in solid-state molecular junctions (MJs) play an important role to modulate or control the electron transport properties. For example, the ions can modulate the conductance of the MJs as theoretically predicted,[1, 2] and observed in several MJs, this conductance modulation depending on the chemical nature of the ions and the details of the molecule-ion conformation.[3-7] Not only the conductance of the MJs, but also the shape of the current-voltage (I-V) behavior can be *in situ* switched between a linear (symmetric) and a rectifying (asymmetric) behavior by inserting/removing the ions or moving the precise position (electrical field driven) of the ions in the MJs,[8-11] opening the way to molecular-scale programmable functional devices and synaptic (neuro-inspired) behavior.[12, 13] Polyoxometalates (POM) are candidates of choice owing to their remarkable reversible multi-redox states[14, 15] and photo-stimulable redox properties,[16, 17] they have attracted a growing interest with several recent demonstration of proof-of-principles as silicon-integrated POM-based memory devices[18] and unconventional (in-memory and neuromorphic) computing devices.[19-21]

In solution, it is well known that the nature of the counterions has a strong impact on the redox potentials for the POMs, hence on the electron transfer rates.[22] In the specific case of POM-based molecular junctions, the interactions with the unavoidable counterions are also controlling many properties of the POM-based materials and devices.[23, 24] For example, for a single POM on a metallic surface, the electron cloud density around the POM, as observed with a scanning tunneling microscope (STM), depends on the counterions that modify the hybridization between the POM and the metal surface.[25] A recent theoretical study shows that the presence of the counterion (whatever its nature, here tetramethylamonium (TMA) vs. Cs$^+$) moves the LUMO (lowest unoccupied molecular orbital) of $W_{18}O_{54}(SO_3)^{2-}$ based MJs closer to the Fermi energy of the electrodes (~0.08 eV above the Fermi energy in both the cases).[26] Albeit this similar position of the LUMO, the computed current-voltage (I-V) curves showed slight differences of the shape (more asymmetric I-V with Cs[+]). The authors suggest that this feature might be due to small differences in the amplitude of the calculated transmission coefficient of electrons in the MJ, clearly calling to more joint experimental and theoretical work. They also note that the presence of the counterions does not create additional new conduction channels in the MJs, but rather that they modify the potential landscape "viewed" by the POMs that transmit the electrons through the junction.

Here, we synthesized a series of Preyssler polyoxometalates (POMs), [Na⊂$P_5W_{30}O_{110}$]$^{14-}$,[27] $P_5W_{30}$ for short, stabilized with four different counterions, H[+], K[+], $NH_4^+$ and tetrabutylammonium ($N(C_4H_9)_4^+$ or TBA[+]). We formed, by electrostatic deposition, monolayers of these POMs on ultraflat Au surface prefunctionalized with a positively charged self-assembled monolayer (SAM) of amine-terminated alkylthiol chains following the protocol already developed and reported in our previous work for other POMs.[17, 28] The electron transport (ET) properties of these POMs were measured by conductive-AFM (atomic force microscopy) and the current vs. voltage curves of these MJs were statistically analyzed (hundreds of I-V traces) to determine the energy position of the molecular orbitals involved in the ET and the electronic coupling between the molecules and the electrodes. We demonstrate that the energy position of the LUMO of the $P_5W_{30}$ with respect of the Fermi energy of the electrodes increases from ~ 0.4 eV to ~ 0.7 eV and that that electrode coupling energy evolves from ~ 0.05 to 1 meV depending on the nature of the counterion in the order K[+], $NH_4^+$, H[+] and TBA[+].

## Synthesis and Structural Characterizations

The synthesis of the $P_5W_{30}$ POMs with various counterions is schematized in the figure 1a. The potassium salt of [Na⊂$P_5W_{30}O_{110}$]$^{14-}$ can be obtained under hydrothermal conditions.[29] The sample we used corresponds to the molecular formula $K_{13.3}Na_{0.7}[Na{\subset}P_5W_{30}O_{110}] \cdot 27H_2O$. The ammonium salt $(NH_4)_{14}[Na{\subset}P_5W_{30}O_{110}] \cdot 25H_2O$ has been prepared according to reference 30. The acid salt $H_{14}[Na{\subset}P_5W_{30}O_{110}] \cdot 44H_2O$ was prepared by ion exchange of a solution of the potassium salt on a Dowex 50W-X8 proton exchange resin, following a published procedure.[31] Acido-basic titration of heteropolyacids with cation-hydroxide is a common route to introduce other cations and it has also been applied to this Preyssler-type phosphotungstic acid.[32] In our hands, the acido-basic titration of $H_{14}[Na{\subset}P_5W_{30}O_{110}]$ with tetrabutylammonium hydroxide was not complete, despite careful colorimetric monitoring of the equivalence. We ended with the mixed salt $TBA_{10}H_4[Na{\subset}P_5W_{30}O_{110}] \cdot 24H_2O$.

We end-up with 4 POMs: $K^+_{13.3}Na^+_{0.7}[Na{\subset}P_5W_{30}O_{110}]^{14-}$ (KP$_5$W$_{30}$ for short); $H^+_{14}[Na{\subset}P_5W_{30}O_{110}]^{14-}$ (HP$_5$W$_{30}$ for short); $TBA^+_{10}H^+_4[Na{\subset}P_5W_{30}O_{110}]^{14-}$ (TBAP$_5$W$_{30}$ for short) and $(NH_4)^+_{14}[Na{\subset}P_5W_{30}O_{110}]^{14-}$ (NH$_4$P$_5$W$_{30}$ for short). See more details, IR and $^{31}$P NMR characterizations, TGA and elemental analysis in the Supporting Information (section 1).

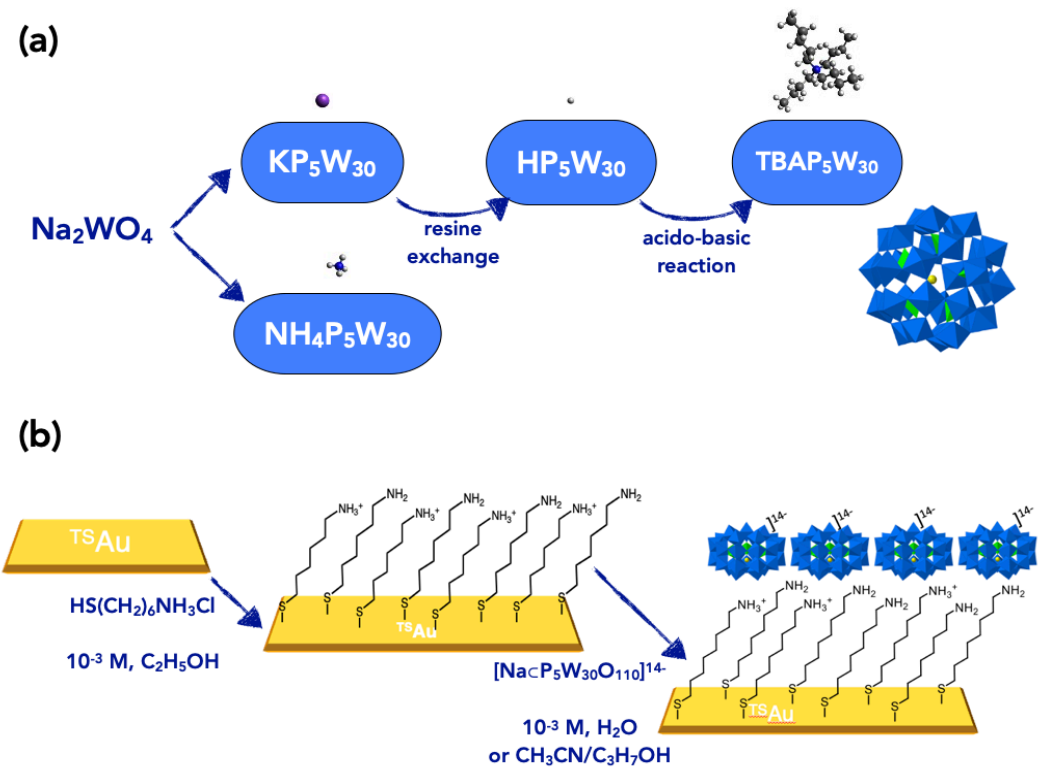

***Figure 1***. *(a) Schematic of the synthesis routes (cations and POM not at scale). (b) Schematic description of the two-step fabrication of the self-assembled monolayers.*

These POMs (dissolved in H$_2$O, or acetonitrile/isopropanol for HP$_5$W$_{30}$ and TBAP$_5$W$_{30}$) were electrostatically deposited on ultra-flat template-stipped $^{TS}$Au substrates functionalized with a 6-aminohexane-1-thiol hydrochloride (HS-(CH$_2$)$_6$-NH$_3^+$/Cl$^-$) self-assembled monolayer (SAM), or C6 SAM for short, following the protocol already developed and reported in our previous work for other POMs (Fig. 1b and also see section 2 in the Supporting Information).[17, 28] The thicknesses (systematically measured by ellipsometry, see section 3 in the Supporting Information) are summarized in Table 1.

| thickness (Å) | KP$_5$W$_{30}$ | HP$_5$W$_{30}$ | NH$_4$P$_5$W$_{30}$ | TBAP$_5$W$_{30}$ |
|---|---|---|---|---|
| C6-SAM | 11±2 | | | |
| POM layer | 9±2 | 10±2 | 9±2 | 14±2 |

*Table 1. Thickness measured by ellipsometry at each step of the fabrication of the molecular junction.*

We note that the P$_5$W$_{30}$ POM has an oval shape with a long axis of 1.8 nm and a short axis of 1.3 nm. Thus the 1.4 nm measured thickness of the TBAP$_5$W$_{30}$ sample indicates a denser monolayer with possibly a small percentage of the TBAP$_5$W$_{30}$ molecule standing upright on the surface (vertical POM orientation), while the thinner values for the 3 other samples can be ascribed to a less compact monolayer whatever the POM orientation (disordered monolayers with various POM orientations). The thicker TBAP$_5$W$_{30}$ SAM may also be caused by the bulky TBA ions (diameter ≈ 0.5-1 nm depending on the conformation of the butyl chains) mainly intercalated between the POMs and the electrode, while small cations can be distributed around the POM.

**Electron Transport Properties**

Figure 2 shows the 2D-histograms (or "heat map") of the I-V curves (between 400 and 500) acquired by C-AFM at different locations on the POM monolayer for the four $^{TS}$Au-C6/POM//Pt MJs ("-" denotes a chemical bond, "/" an electrostatic contact and "//" a mechanical contact). The main feature is that these I-V distributions have almost the same level of currents at large voltages, i.e. |V|>1V, while the currents in the voltage range [-0.6 V ; 0.6 V] gradually increase with changing the counterion in the order from K$^+$, NH$_4$$^+$, H$^+$ to TBA$^+$. This is clearly seen in Fig. 3a, plotting the mean current - voltage ($\bar{I}$-V) for the 4 samples and their counterion dependence at several voltages (Fig. 3b). The mean current $\bar{I}$ is around 10$^{-8}$ A at +/- 1.5V, while it varies over 2 decades at +/- 0.5 V. This observation is also supported by plotting and comparing the current histograms



at given voltages (here at +/- 0.5V and +/- 1V), Figs. S9 and S10 in the Supporting Information.

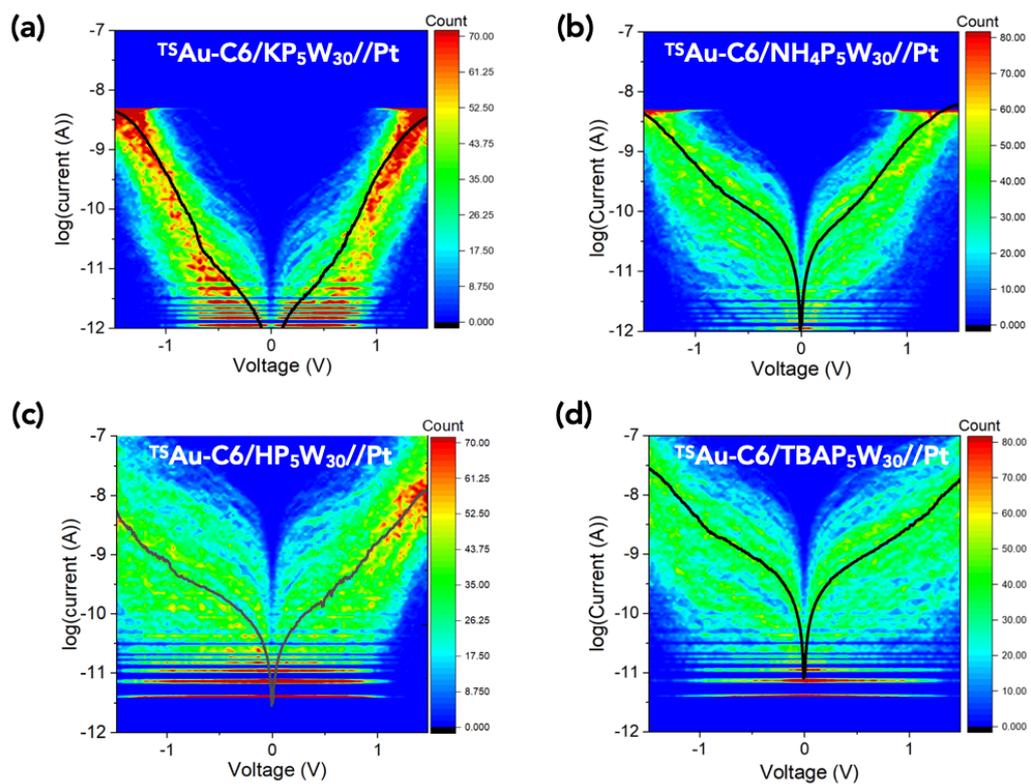

*Figure 2*. 2D histograms (400-500 I-V traces) of the four samples in a semi-log |I|-V plot. The solid black lines are the calculated mean Ī-V curves.



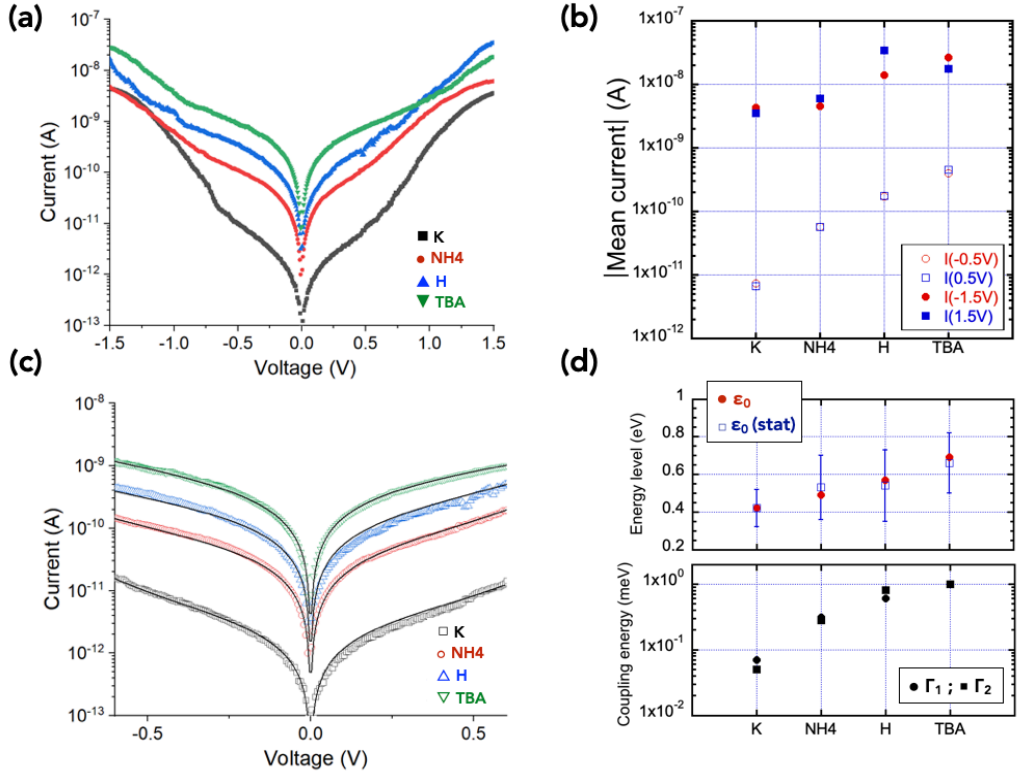

*Figure 3.* (a) Mean Ī-V curves for the four samples (from the datasets shown in Fig. 3), (b) counterion dependence of the mean current Ī at +/- 1.5 V and +/- 0.5 V, (c) fits (solid lines) of the SEL model in the voltage window [-0.6 V ; 0.6 V] (fit $R^2$= 0.995 for $K^+$, 0.998 for $NH_4^+$, 0.995 for $H^+$, 0.999 for $TBA^+$), (d) counterion evolution of the SEL model parameters $\varepsilon_0$, $\Gamma_1$ and $\Gamma_2$.

The Ī-V shapes for all samples show a bump at large voltages, i.e. |V|>0.7 - 1V, while the low-voltage Ī-Vs are well fitted between -0.6 and 0.6 V with a single-energy level (SEL) model given by the following analytical expression:[33, 34]

$$I(V) = N\frac{8e}{h}\frac{\Gamma_1\Gamma_2}{\Gamma_1+\Gamma_2}\left[\arctan\left(\frac{\varepsilon_0 + \frac{\Gamma_1}{\Gamma_1+\Gamma_2}eV}{\Gamma_1+\Gamma_2}\right) - \arctan\left(\frac{\varepsilon_0 - \frac{\Gamma_2}{\Gamma_1+\Gamma_2}eV}{\Gamma_1+\Gamma_2}\right)\right]$$

(1)



with $\varepsilon_0$ the energy of the molecular orbital (MO), here LUMO involved in the transport (with respect to the Fermi energy of the electrodes), $\Gamma_1$ and $\Gamma_2$ the electronic coupling energy between the MO and the electron clouds in the two electrodes, e the elementary electron charge, h the Planck constant and N the number of molecules contributing to the ET in the molecular junction (see details in section 6 the Supporting Information). Among several limitations (section 6 in the Supporting Information), the SEL model is a low temperature approximation and the temperature broadening of the Fermi function in the electrodes is not taken into account. However, it was shown that it can be reasonably used at room temperature for voltages below the sharpened increase of the current (not observed here in the -0.6/0.6 V window) characterizing the transition between the off-resonance and resonant transport conditions at which the broadening of the Fermi function modify the I-V shape.[35-37]

Fig 3c shows typical examples of the SEL fits on the mean $\bar{I}$-V and Fig. 3d shows the evolution of the fitted parameters versus the nature of the counterions, clearly revealing an increase of the molecular energy level $\varepsilon_0$ from 0.42 eV (for $KP_5W_{30}$) to 0.69 eV (for $TBAP_5W_{30}$). The electrode coupling energies, $\Gamma_1$ and $\Gamma_2$, follow a same trend, increasing from ~0.08 to 1-2 meV in the same counterion order (Fig. 3d and Table 2). The same trend is confirmed from a statistical analysis by fitting with the SEL model all the individuals I-Vs of the datasets shown in figure 2. The statistical distributions of $\varepsilon_0$, $\Gamma_1$ and $\Gamma_2$ for the 4 MJs are shown in Figs. S12 and S13 (Supporting Information) and the mean values of these distributions are summarized in Fig. 3d and Table 2.

The same results are obtained by analyzing the I-V curves with the transition voltage spectroscopy (TVS) method.[38-42] The energy level and the electrode coupling energy determined by the fit of the SEL model and by the TVS method are in very good agreement (section 7, Figs. S14 and S15 in the Supporting Information).



Albeit the statistical analysis shows a large distribution of values (e.g. 2 decades for the parameters $\Gamma_1$ and $\Gamma_2$ - Fig. S13 in the Supporting Information, and Table 2), the same trends and order of magnitude of these parameters are observed from the analysis of the mean Ī-V, the statistical analysis of the full dataset, and using both the SEL and TVS methods, making the experimental observations conclusive.

|  | $\varepsilon_0$ (eV) | $\Gamma_1$ (meV) | $\Gamma_2$ (meV) | ⟨$\varepsilon_0$⟩ (eV) | ⟨$\Gamma_1$⟩ (meV) | ⟨$\Gamma_2$⟩ (meV) |
|---|---|---|---|---|---|---|
| KP$_5$W$_{30}$ | 0.42 | 0.085 | 0.079 | 0.42±0.10 | 0.07 (0.025/0.2) | 0.05 (0.02/0.13) |
| NH$_4$P$_5$W$_{30}$ | 0.49 | 0.33 | 0.39 | 0.53±0.17 | 0.31 (0.1/1) | 0.28 (0.09/0.83) |
| HP$_5$W$_{30}$ | 0.57 | 0.68 | 0.81 | 0.54±0.19 | 0.61 (0.15/2.4) | 0.51 (0.14/1.86) |
| TBAP$_5$W$_{30}$ | 0.69 | 2.0 | 1.0 | 0.66±0.16 | 1 (0.24/4.3) | 0.99 (0.25/3.8) |

*Table 2.* SEL parameters $\varepsilon_0$, $\Gamma_1$ and $\Gamma_2$ fitted on the mean Ī-V (Fig. 3a) and mean values ⟨$\varepsilon_0$⟩, ⟨$\Gamma_1$⟩ and ⟨$\Gamma_2$⟩ deduced from the statistical analysis (see text). For the energy level ⟨$\varepsilon_0$⟩, the error is the standard deviation of the normal distribution (Fig. S12). For the ⟨$\Gamma_1$⟩ and ⟨$\Gamma_2$⟩ parameters, the values in parenthesis are the min and max values at FWHM of the log-normal distributions (Fig. S13).

## Discussion

The low-bias conductance of the P$_5$W$_{30}$-based MJs increases by a factor ~ 100 between the K$^+$ and the TBA$^+$ cations (Figs. 3b-c). However, the weak increase of the energy position of the LUMO (factor ~ 1.6, Fig. 3d, Table 2) should have induced a decrease of the current due to a higher electron energy barrier ($\varepsilon_0$-$\varepsilon_F$) at the molecule/electrode interface. A factor ~ 4.5 is estimated from Eq. (1) assuming the same molecule/electrode coupling energy for simplicity ($\Gamma_1$=$\Gamma_2$=0.1



meV) in the simulated I-V curves (Fig. S16 in the Supporting Information). The increase of the measured current is clearly related to the larger increase (factor ~ 20, Fig. 3d and Table 2) of the molecule/electrode coupling energy. We examine several hypotheses to explain these results. The trend observed in solution for the size-depending behavior of the alkali salts of POMs is barely transposable: indeed, a positive shift of the reduction potential (which would correspond here to a stabilization of the LUMO and a decrease of $\varepsilon_0$) and a greater electron transfer rate to POMs associated with larger alkali metal cations had been ascribed to an increase of the ion pairing energy related to a decrease of the cation solvation and the formation of more intimate cation/anion pairs.[22] Although we cannot completely exclude the presence of some remaining solvent molecules in our MJs, solvation is not really relevant here. In a recent theoretical report, the efficiency of cations to compensate for the charging of a hexavanadate POM has been addressed by simultaneous addition of extra cations and electrons: in the gas phase, $Li^+$ and $H^+$ were found the most efficient, followed by $NH_4^+$ and $K^+$.[43] Although the origin of this effect was not further discussed it could similarly impact the electron transport across the MJs. Last but not least, the tetrabutylammonium cation is very different from the other cations used in this study, and because of its volume and lower charge density the softer cation of the series.

Among many factors, the energy level alignment in the MJs (LUMO of the $P_5W_{30}$ POM with respect to the electrode Fermi energy) may depend on the interface dipole. The larger TBA counterion could probably induce the largest local dipole (larger distance between the positive and negative charges), and if we assume that the counterions are mostly inserted between the POM monolayer and the electrode (due to steric hindrance, *vide supra* the thickness measurements), then a global dipole could exist at the interface with an orientation leading to an upwards larger LUMO offset from the Fermi level due to the dipole-induced vacuum level shift (Fig. S17 in the Supporting Information).



On the contrary, smaller cations create weaker dipoles randomly arranged around the POM,[44] resulting in a weaker (or even negligible) average dipole at the interface, thus a weaker LUMO shift. Note that we cannot exclude, at this stage without further theoretical explorations, that the atomistic details of the POM/electrode contact configuration play also a role (as also known in many other MJs). In the case of the $W_{18}O_{54}(SO_3)^{2-}$ based MJs,[26] a change by a factor ca. 4 in the amplitude of the current and a variation of the LUMO between ca. 0.4 and 1.2 eV was theoretically predicted whether the POM is oriented with its long axis perpendicular or horizontal between the electrodes and how (atomic configuration) the POM is connected to the Au electrode. From the thickness measurements (*vide supra*), we also suggest possible different orientation of the POMs, thus further detailed calculations (out of the scope of this work) are called to a reach a conclusion.

The evolution of the molecule/electrode coupling energy is most intriguing. Nevertheless, it has been recently demonstrated that the counterion can mediate the intermolecular electron transfer between two adjacent and nearby POMs,[45] leading to a possible POM-cation-POM electron conduction channel. In the situation schematically depicted above for the case of the $TBA^+$ counterion (Fig. S17 in the Supporting Information), we extrapolate that the same kind of mechanism could enhance the POM-cation-electrode electronic coupling, compared to a more random organization around the POMs for the other counterions (Fig. S17). Moreover, the fact that the POM/electrode interaction also depends on the nature of the counterions has been observed by STM images of a single POM deposited on a metal surface. In this experiment, the size of the electron cloud density measured by STM largely extends the geometric size of the POM due to the hybridization of the POM and the counterion with the metal surface, with a stronger hybridization with a $TBA^+$ cation rather than the POM encapsulated in a cyclodextrin cage (cation free).[25] This hybridization is also the key factor fixing the molecule/electrode energy



coupling in the SEL model (Eq. 1). Thus, we hypothesis that the $TBA^+$ counterion could lead to a stronger POM/electrode hybridization than the other ones. Finally, we note that our energy values with the $TBA^+$ counterion ($\varepsilon_0$ = 0.66-0.69 eV, $\Gamma_1$ and $\Gamma_2 \approx$ 1-2 meV) are of the same order as values determined by STM experiments on a single similar Preyssler POM ($[DyP_5W_{30}O_{110}]^{12-}$) in the case of a strong molecule/electrode coupling (established by moving the STM tip in close contact to the molecule in that case): $\varepsilon_0 \approx$ 0.7 eV, $\Gamma_1$ and $\Gamma_2 \approx$ 1-10 meV.[46]

At higher voltages ($|V| > 0.7$ V), the bump of the current cannot be fitted with the SEL model. We hypothesis that this "additional" current can be due to a second level entering the energy window at higher voltages (Fig. S11 in the Supporting Information). Such a behavior was theoretically simulated in the case of a two-channel electron transmission in the MJs.[47] This second level is tentatively attributed to the LUMO+1 of the POMs. Our results suggest that the LUMO+1 is less sensitive to the counterions, since the currents at $|V| > 0.7$ V are less dependent on them (Fig. 3b and Fig. S10 in the Supplementary Information). We also note that the POM samples with the $NH_4^+$ and $H^+$ countercations display a slightly higher current at +1V than at -1V (Fig. S10 and Table S1 in the Supporting Information), while no voltage dependence asymmetry is observed at lower voltages (Fig. S9). The mean asymmetry ratio (also known as rectification ratio) $R=\bar{I}(+1V)/\bar{I}(-1V)$ is weak ($\approx$1.7 and $\approx$3.2 for $HP_5W_{30}$ and $NH_4P_5W_{30}$, respectively) with a difference in log-$\bar{I}$ (log-mean current) of 0.24 for $HP_5W_{30}$ and 0.5 for $NH_4P_5W_{30}$, smaller than log-$\sigma$ (log standard deviation, in the range 0.84-1.1). It has been pointed out that weak R values must be considered with caution without a solid statistical analysis.[48] A two-sample t-test on the current datasets at +1 and -1V was used to assess if these datasets statistically and significantly differ from each other (see section 9 in the Supporting Information). The t-test results show that the null hypothesis (same mean value for the two datasets) can be rejected and this weak current asymmetry is statistically significant. The origin of this effect may be due to different contact geometries of



the POM to the electrodes (at the atomic level) in the presence of the different conterions, with an asymmetric hybridization between the POM and the two electrodes for the LUMO+1 in the case of $H^+$ and $NH_4^+$ counterions (no asymmetry is observed at lower bias, -0.6/0.6V, involving only the LUMO in the electron transport mechanism (Fig. 3c and Fig. S9 in the Supporting Information). A similar effect was theoretically predicted for $W_{18}O_{54}(SO_3)^{2-}$ POM with TMA (symmetric I-V) or $Cs^+$ (asymmetric I-V) counterions.[26] Further theoretical work (out of the scope of this work) would be necessary to clarify this point.

## Conclusion

In conclusion, we have demonstrated that the electron transport properties, at the nanoscale, of $P_5W_{30}$-based molecular junctions depend on the nature of the counterions. From a statistical study (hundreds of I-V traces) using a simple analytical model derived of the Landauer-Imry-Büttiker formalism for charge transport in nanoscale devices, we have found that the energy position of the LUMO of the $P_5W_{30}$ with respect of the Fermi energy of the electrodes increases from ~ 0.4 eV to ~ 0.7 eV and that that electrode coupling energy evolves from ~ 0.05 to 1 meV depending on the nature of the counterion in the order from $K^+$, $NH_4^+$, $H^+$ to $TBA^+$. We have suggested that these variations could be due to a counterion-dependent POM/electrode interface dipole and a molecule/electrode hybridization mediated by the counterions. These non-trivial POM-counterion-metal electrode interactions clearly require further theoretical studies to elucidate the influence of the counterions on the electron transport properties of POM-based molecular devices.

## Methods

***Sample fabrication.***



*Bottom metal electrode fabrication.* Template stripped gold (<sup>TS</sup>Au) substrates were prepared according to the method previously reported.[49-51] (section 2 in the Supporting Information).

*Self-assembled monolayers*. The SAMs on <sup>TS</sup>Au were fabricated following a protocol developed and optimized in our previous works for the electrostatic immobilization of POMs on amine-terminated SAMs.[17, 28] (section 2 in the Supporting Information).

**Spectroscopic ellipsometry.**

The thickness of the SAMs was measured by spectroscopic ellipsometry (UVISEL ellipsometer (HORIBA), section 3 in the Supporting Information)

**CAFM in ambient conditions.** We measured the electron transport properties at the nanoscale by CAFM (ICON, Bruker) at room temperature using a tip probe in platinum/iridium (see section 4 in the Supporting Information).

## Associated content

The Supporting Information is available free of charge at xxxxxx.

Synthesis details, IR and NMR characterizations; fabrication of self-assembled monolayers; ellipsometry measurements; protocol of C-AFM measurements; supplemental details on the I-V measurements (statistical data); details on the single energy level (SEL) model and on the statistical distribution of the model paramaters; transition voltage spectroscopy (details on the technique and results); supplementary figures for the discussion.

*Author Contributions*

C.H., S.R., F.V. and A.P. synthesized and characterized the POMs and fabricated the SAMs. C.H. did all the C-AFM measurements, C.H. and D.V. analyzed the data. A.P. and D.V. conceived and supervised the project. The manuscript was written by D.V. with the contributions of all the authors. All authors have given approval of the final version of the manuscript.




***Note***

The authors declare no competing financial interest.

## Acknowledgements.

We acknowledge support of the CNRS (France), project neuroPOM, under a grant of the 80PRIME program. We are thankful to Dr. Masahiro Sadakane for sharing his knowledge about the preparation of the tetrabutylammonium salt of the Preyssler anion.

# Experimental observation of the role of countercations on the electrical conductance of Preyssler-type polyoxometalate nanodevices.


Cécile Huez,[1,‡] Séverine Renaudineau,[2] Florence Volatron,[2]
Anna Proust[2] and Dominique Vuillaume.[1]

1) Institute for Electronics Microelectronics and Nanotechnology (IEMN), CNRS, University of Lille, Av. Poincaré, Villeneuve d'Ascq, France
2) Institut Parisien de Chimie Moléculaire (IPCM), CNRS, Sorbonne Université, 4 Place Jussieu, Paris, France
‡) Now at : Instituto de Ciencia Molecular (ICMoL), University of Valencia, Spain.


**SUPPORTING INFORMATION**

1. **Synthesis Details**

All the chemicals were used as supplied. $K_{12,5}Na_{1,5}[Na \subset P_5W_{30}O_{110}]$,[1] $(NH_4)_{14}[Na \subset P_5W_{30}O_{110}]$,[2] and $H_{14}[Na \subset P_5W_{30}O_{110}]$,[3] were prepared according to published procedures and their purity checked by IR and $^{31}P$ NMR spectroscopy. The synthesis of $TBA_{10}H_4[Na \subset P_5W_{30}O_{110}]$ was adapted from the literature.[4]

IR spectra (KBr pellets) have been recorded in transmission from 250 to 4000 $cm^{-1}$ on a Jasco FT/IR 4100 spectrometer (Resolution 4 $cm^{-1}$). $^{31}P$ NMR spectra (121 MHz) have been recorded on a Bruker Avance II 300MHz spectrometer and the chemical shifts referenced to external 85 % $H_3PO_4$. Thermogravimetric

analyses (TGA) have been performed on SDT-Q600 TA Instrument thermobalance under an air flow to assess the amount of water molecules. Elemental analyses have been carried out at Crealins and at the Institut des Sciences Analytiques, CNRS, Villeurbanne, France.

### Synthesis and characterization of $K_{12,5}Na_{1,5}[Na{\subset}P_5W_{30}O_{110}]$

To 33g of $Na_2WO_4.2H_2O$ dissolved in 30mL of water were added 26.5mL of $H_3PO_4$ (85%). This mixture was placed in a solvothermal synthesis bomb (Parr Model 4748), heated to 120°C overnight. Once the yellow solution has been cooled to room temperature, 15mL of $H_2O$ was added followed by 10g of KCl. The precipitate was membrane filtered and washed with 2M potassium acetate and methanol. Recrystallization in hot water was necessary to purify the product (with a pause time < 2h to avoid the formation of crystals of $[a-P_2W_{18}O_{62}]^{6-}$). About 8g of crystals are then collected (yield about 30%). A second recrystallization can be performed if the product still contains impurities. For each step, the purity was assessed by $^{31}P$ NMR: the small signal at -12,55 ppm corresponds to $[a-P_2W_{18}O_{62}]^{6-}$.[5] Thermogravimetric analysis gave 27 water molecules of crystallization.

IR (KBr, cm$^{-1}$): n 3538 (m), 3432 (m), 1616 (m), 1164 (m), 1082 (m), 1017 (w), 982 (sh), 936 (s), 911 (s), 782 (s), 670 (sh), 571 (w), 538 (w), 464 (w), 349 (m), 311 (w), 286 (w). $^{31}P$ NMR (D$_2$O): δ = -9.61 ppm. Elemental analysis: found (calculated for $K_{13,3}Na_{0,7}[Na{\subset}P_5W_{30}O_{110}].27H_2O$ MW = 8475.07 g mol$^{-1}$) K: 6.12 (6.14), Na: 0.42 (0.46), P: 1.76 (1.83), W: 65.23 (65.08) %.



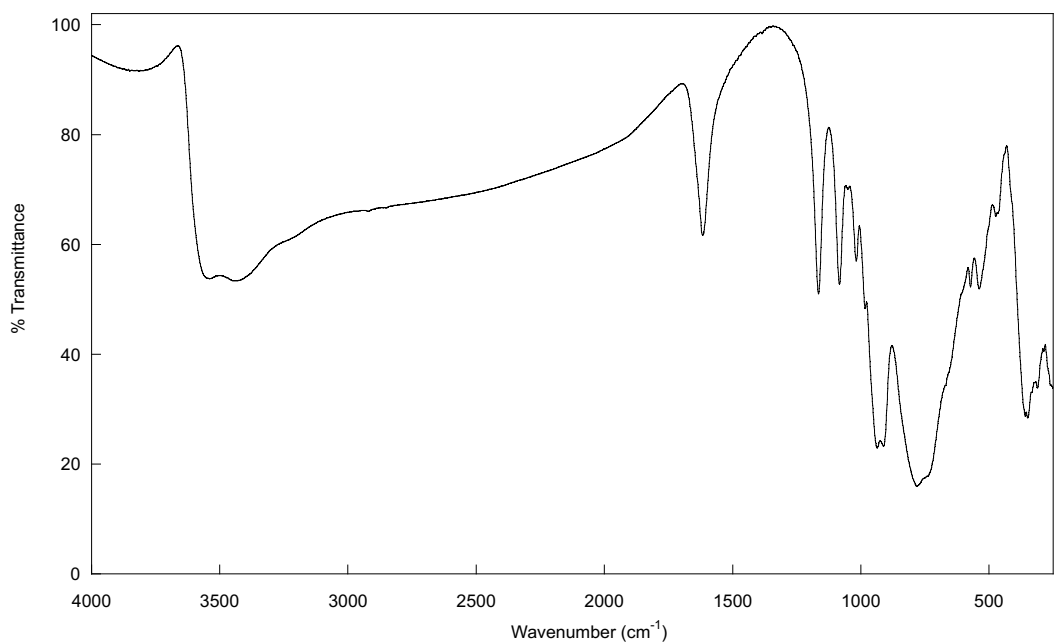

*Figure S1*. IR spectrum of $K_{13.3}Na_{0.7}[Na \subset P_5W_{30}O_{110}]$ (KBr pellet).

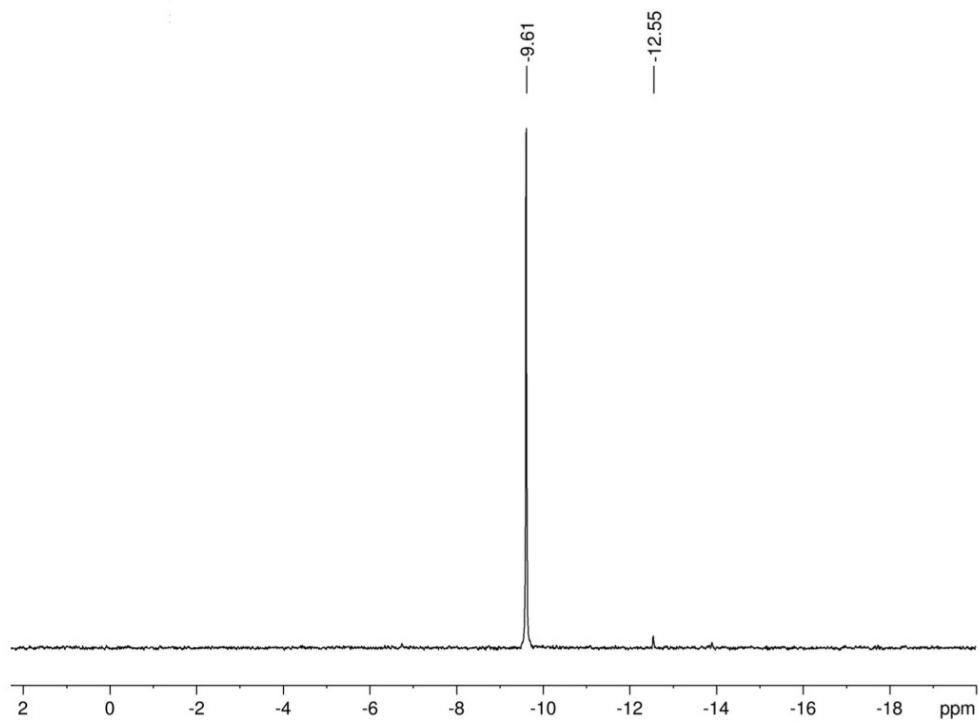

*Figure S2.* $^{31}P$ NMR spectrum of $K_{13.3}Na_{0.7}[Na \subset P_5W_{30}O_{110}]$ in $D_2O$.



**Synthesis of $(NH_4)_{14}[Na \subset P_5W_{30}O_{110}]$**

The synthesis was performed following the published procedure and thermogravimetric analysis carried out on an air-dried sample gave 25 water molecules of crystallization.

IR (KBr, cm$^{-1}$): n 3551 (m), 3443 (m), 3127 (m), 1615 (m), 1400 (s), 1165 (m), 1080 (m), 1017 (m), 982 (w), 934 (s), 910 (s), 777 (s), 571 (w), 539 (w), 470 (w), 359 (m), 311 (w). $^{31}$P NMR (D$_2$O): δ = -9.61 ppm.

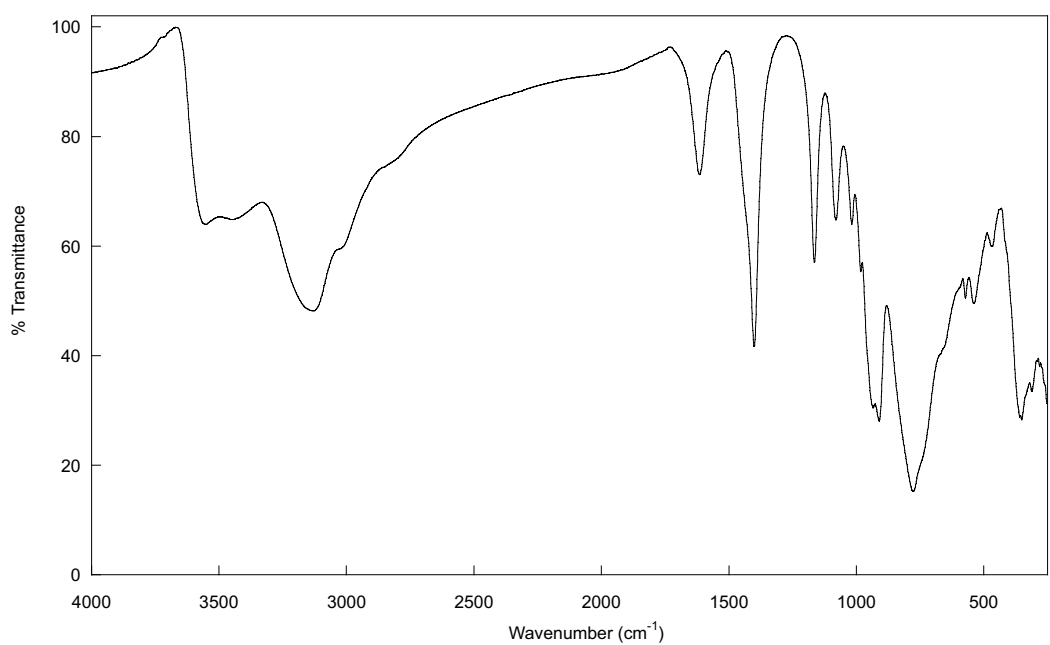

*Figure S3*. IR spectrum of $(NH_4)_{14}[Na \subset P_5W_{30}O_{110}]$ (KBr pellet).



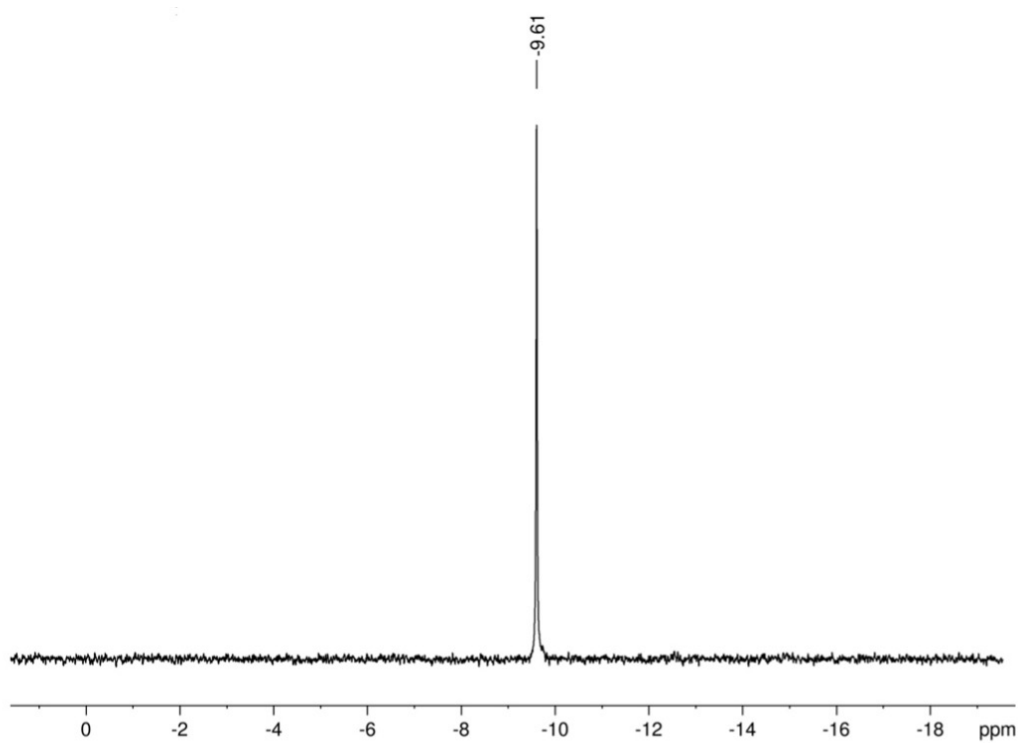

***Figure S4***. *31P NMR spectrum of (NH$_4$)$_{14}$[Na⊂P$_5$W$_{30}$O$_{110}$] in D$_2$O.*

**Synthesis of H$_{14}$[Na⊂P$_5$W$_{30}$O$_{110}$]**

First, we prepared the Dowex column 50W-X8 in its acidic form. In a beaker, the very fine particles of the resin were separated from the mass by successive sedimentations and settlings in distilled water. The resin was then washed with distilled water until the orange coloration resulting from the degradation of the polymer disappeared. Finally, the resin was placed in a column (diameter 1.5 cm, length 40 cm), washed three times with its dead volume of 1M HCl (~3x20 mL) and rinsed with water until the pH obtained for the eluent reached that of distilled water.[6] In a second step, 8.2g of K$_{13.3}$Na$_{0.7}$[Na⊂P$_5$W$_{30}$O$_{110}$].27H$_2$O were dissolved in 160 mL of H$_2$O. The solution was transferred to the column and passed through the resin at a rate of one drop per second. Water (~40mL) was then passed through the column, as long as the pH of the eluent remained acidic to ensure that all the product was collected. The eluent was then evaporated at a



temperature of 60-65°C to recover the product as a yellow and sticky solid, which was crushed twice with ethanol, then dried with diethylether (yield 7.8g). According to thermogravimetric analysis, 44 water molecules are present leading to the general formula $H_{14}[Na \subset P_5W_{30}O_{110}] \cdot 44H_2O$.

IR (KBr, cm$^{-1}$): n 3400 (m), 3199 (w), 1708 (w), 1624 (m), 1604 (m), 1163 (m), 1080 (m), 1022 (m), 939 (s), 914 (s), 765 (s), 670 (sh), 571 (w), 532 (w), 465 (w), 355 (m), 342 (m), 301 (w). $^{31}$P NMR (D$_2$O): δ = -9.61 ppm. Elemental analysis: found (calculated for $H_{14}[Na \subset P_5W_{30}O_{110}] \cdot 44H_2O$ M = 8260.08 g mol$^{-1}$) H: 1.47 (1.24), Na: 0.38 (0.28), P: 1.68 (1.87), W: 66.34 (66.77) %.

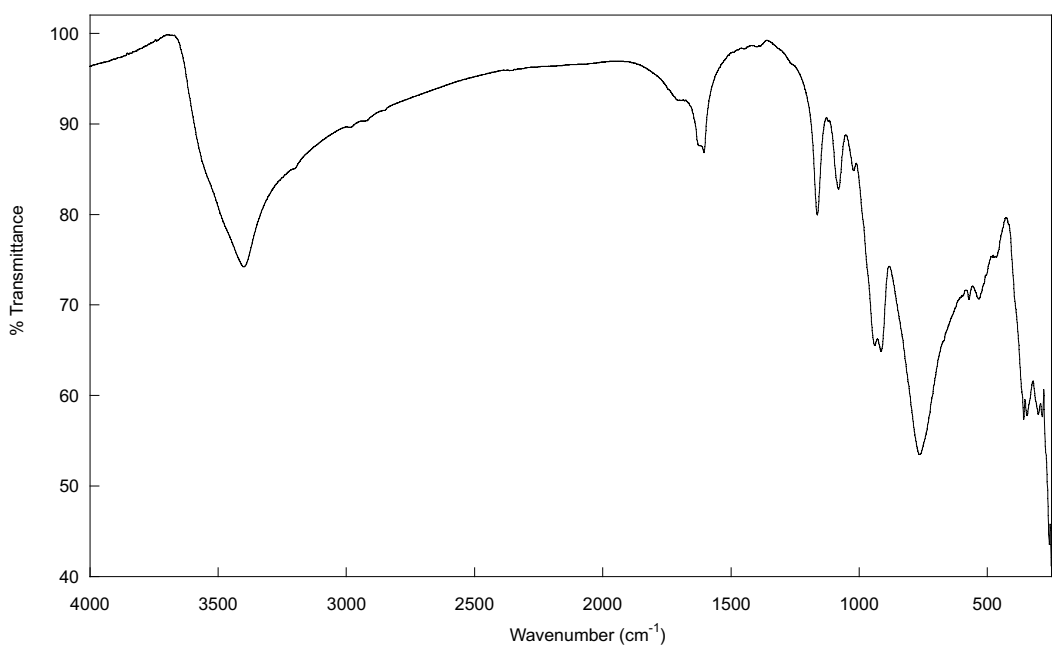

*Figure S5. IR spectrum of $H_{14}[Na \subset P_5W_{30}O_{110}]$ (KBr pellet).*



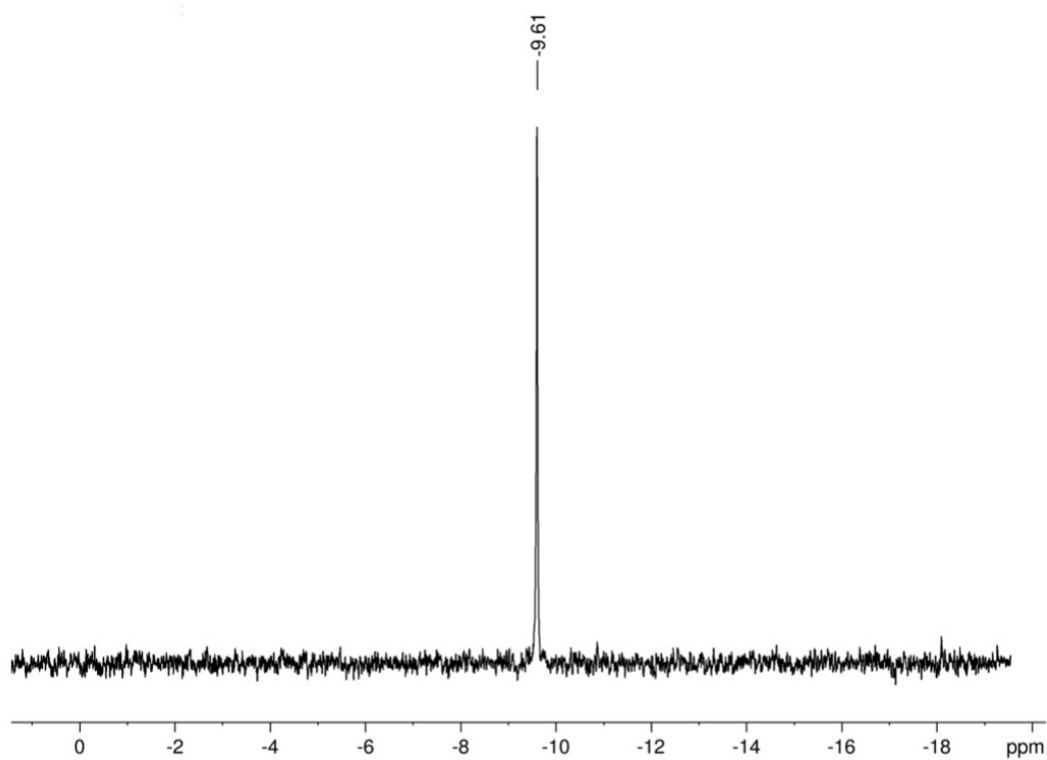

*Figure S6*. $^{31}$P NMR spectrum of $H_{14}[Na \subset P_5W_{30}O_{110}]$ in $D_2O$.

**Synthesis of TBA$_{10}$H$_4$[Na⊂P$_5$W$_{30}$O$_{110}$]** adapted from reference 3.

In a bicol flask, 6.0 g (0.73 mmole) of $H_{14}[Na \subset P_5W_{30}O_{110}] \cdot 44H_2O$ were dissolved in 100 mL of degassed distilled water, 2 drops of phenolphthalein were added to clearly see the turning point. The solution was titrated with a 1.54 M TBAOH solution (40 wt% in water, independently titrated) and the titration was monitored with a pH electrode under a nitrogen flow. The TBAOH solution was added slowly with a syringe, the solution became milky very quickly but cleared again shortly before the turning point. After addition of 4.8 mL of the 1.54 M TBAOH solution plus 0.1 mL of a 5-fold diluted (0.308 M) TBAOH solution, the solution remained slightly pinkish. It was then evaporated under vacuum. During the evaporation, a fading of the solution was observed which showed that the pH slightly decreased, the evaporation was stopped for addition of a small amount of the TBAOH solution. The phenomenon was repeated 3 times bringing the



equivalent number of TBAOH to 7.70 mmoles (instead of the 10.22 mmoles expected for titration of the 14 protons per polyoxometalate). The product was dried with 2x10 mL absolute ethanol and 10 mL ether. The presence of 24 water molecules was determined by thermogravimetric analysis, leading to the formula TBA$_{10}$H$_4$[Na⊂P$_5$W$_{30}$O$_{110}$].24H$_2$O.

IR (KBr, cm$^{-1}$): n 3407 (m), 2960 (m), 2933 (w), 2871 (m), 2741 (w), 1636 (m), 1484 (m), 1380 (w), 1160 (m), 1075 (m), 1018 (m), 982 (m), 918 (s), 794 (s), 747 (s), 573 (w), 541 (w), 474 (w), 369 (m), 312 (m). $^{31}$P NMR (CD$_3$CN) δ = -9.06 ppm. Elemental analysis: found (calculated for TBA$_{10}$H$_4$[Na⊂P$_5$W$_{30}$O$_{110}$].24H$_2$O MW = 10314.41 g.mol$^{-1}$) C: 19.26 (18.63) H: 3.96 (4.03), N: 1.45 (1.36); Na: 0.32 (0.22), P: 0.91 (1.50), W: 52.66 (53.47)%.

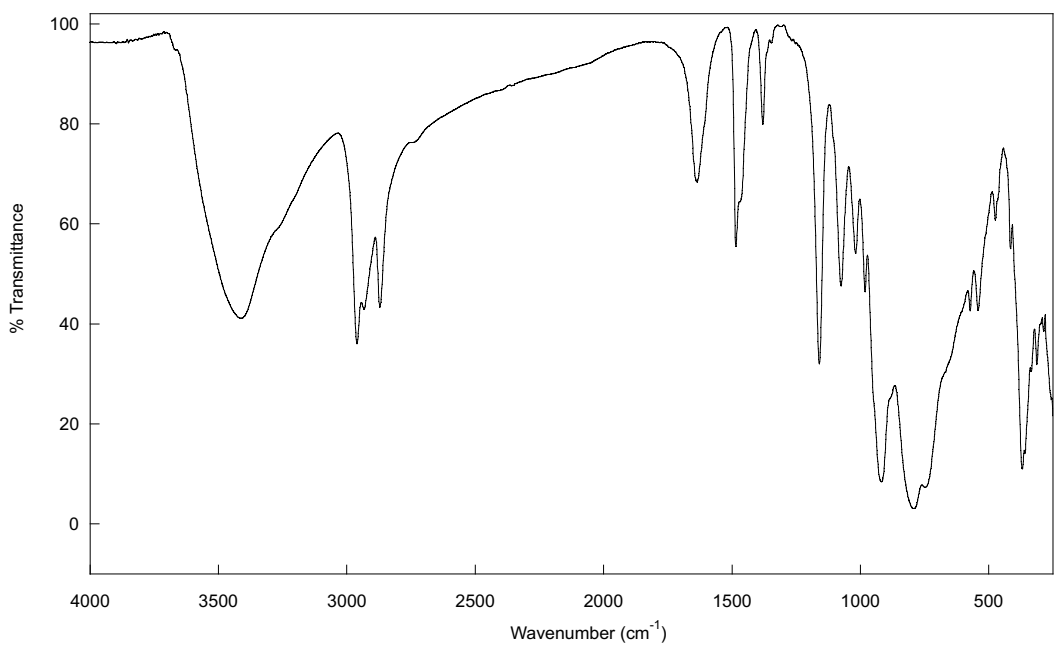

*Figure S7*. IR spectrum of TBA$_{10}$H$_4$[Na⊂P$_5$W$_{30}$O$_{110}$] (KBr pellet).



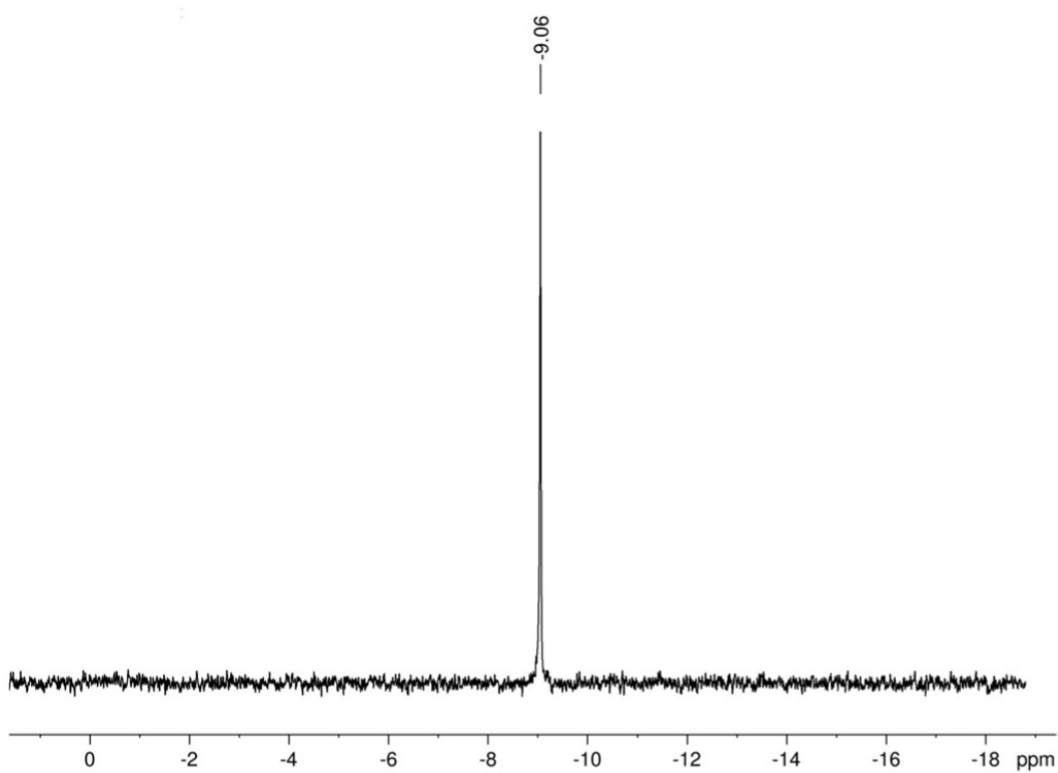

**Figure S8**. $^{31}P$ NMR spectrum of $TBA_{10}H_4[Na{\subset}P_5W_{30}O_{110}]$ in $CD_3CN$.

## 2. Self-Assembled Monolayer

Ultraflat template-stripped gold surfaces ($^{TS}$Au), with rms roughness of ~0.4 nm were prepared according to the method already reported.[7-9] In brief, a 300–500 nm thick Au film was evaporated on a very flat silicon wafer covered by its native $SiO_2$ (rms roughness of ~0.4 nm), which was previously carefully cleaned by piranha solution (30 min in 7:3 $H_2SO_4/H_2O_2$ (v/v); **Caution**: Piranha solution is a strong oxidizer and reacts exothermically with organics), rinsed with deionized (DI) water, and dried under a stream of nitrogen. Clean 10x10 mm pieces of glass slide (ultrasonicated in acetone for 5 min, ultrasonicated in 2-propanol for 5 min, and UV irradiated in ozone for 10 min) were glued on the evaporated Au film (UV-polymerizable glue, NOA61 from Epotecny), then mechanically peeled off



providing the $^{TS}$Au film attached on the glass side (Au film is cut with a razor blade around the glass piece).

The self-assembled monolayers (SAMs) of 6-aminohexane-1-thiol hydrochloride (HS-$(CH_2)_6$-$NH_3^+$/ $Cl^-$) were prepared following a protocol optimized and described in a previous work for the electrostatic immobilization of POMs on amine-terminated SAMs.[10, 11] The freshly prepared $^{TS}$Au substrates were dipped in a solution of 6-aminohexane-1-thiol hydrochloride (Sigma-Aldrich) at a concentration of $10^{-3}$ M in ethanol overnight in the dark. The samples were rinsed in ethanol for 5 min and then ultrasonically cleaned 5 min in deionized (DI) water. These SAMs were treated by a PBS (phosphate-buffered saline, pH=7.4) solution for 2 hours, followed by ultra-sonication in DI water for 5 minutes. The substrates were finally washed with ethanol and dried under nitrogen flow. It was found that the PBS treatment removes the formation of aggregates on the aminoalkylthiol SAMs as well as avoids clustering of POMs during the electrostatic deposition, likely because this treatment optimizes the ratio of $NH_3^+$/$NH_2$ on the surface.[10] The electrostatic deposition of the POMs was done by immersion of these SAMs in a solution of POM at a concentration of $10^{-3}$ M in $H_2O$, except for $HP_5W_{30}$ and $TBAP_5W_{30}$ dissolved in ACN/isopropanol, for one to few hours. We checked by ellipsometry that the thickness of the POM layer was independent of the immersion time when the immersion time was longer than 1h.

3. **Ellipsometry Measurements**

We recorded spectroscopic ellipsometry data (on *ca.* 1 cm$^2$ samples) in the visible range using a UVISEL (Horiba Jobin Yvon) spectroscopic ellipsometer equipped with DeltaPsi 2 data analysis software. The system acquired a spectrum ranging from 2 to 4.5 eV (corresponding to 300–750 nm) with intervals of 0.1 eV (or 15 nm). The data were taken at an angle of incidence of 70°, and the compensator



was set at 45°. We fit the data by a regression analysis to a film-on-substrate model as described by their thickness and their complex refractive indexes. First, a background for the substrate before monolayer deposition was recorded. We acquired three reference spectra at three different places of the surface spaced of few mm. Secondly, after the monolayer deposition, we acquired once again three spectra at three different places of the surface and we used a 2-layer model (substrate/SAM) to fit the measured data and to determine the SAM thickness. We employed the previously measured optical properties of the substrate (background), and we fixed the refractive index of the monolayer at 1.50.[12] We note that a change from 1.50 to 1.55 would result in less than a 1 Å error for a thickness less than 30 Å. The three spectra measured on the sample were fitted separately using each of the three reference spectra, giving nine values for the SAM thickness. We calculated the mean value from this nine thickness values and the thickness incertitude corresponding to the standard deviation. Overall, we estimated the accuracy of the SAM thickness measurements at ± 2 Å.[13]

**4. C-AFM Measurements**

We measured the electron transport properties at the nanoscale by CAFM (ICON, Bruker) at room temperature using a tip probe in platinum/iridium. We used a "blind" mode to measure the current-voltage (I-V) curves and the current histograms: a square grid of 10×10 was defined with a pitch of 50 to 100 nm. At each point, the I-V curve is acquired leading to the measurements of 100 traces per grid. This process was repeated several times at different places (randomly chosen) on the sample, and up to several thousand of I-V traces were used to construct the current-voltage histograms (shown in Fig. 3, main text).

The tip load force was set at ≈6-9 nN for all the I-V measurements, a lower value leading to too many contact instabilities during the I-V measurements. Albeit larger than the usual load force (2-5 nN) used for CAFM on SAMs, this value is



below the limit of about 60-70 nN at which the SAMs start to suffer from severe degradations. For example, a detailed study (Ref. 14) showed a limited strain-induced deformation of the monolayer (≲ 0.3 nm) at this used load force. The same conclusion was confirmed by our own study comparing mechanical and electrical properties of alkylthiol SAMs on flat Au surfaces and tiny Au nanodots.[15]

## 5. Details of the I-V measurements

From the datasets shown in Fig. 2, we extracted the distribution of the current measured at +/- 0.5V (Fig. S9) and +/- 1V (Fig. S10) to support the results shown in Fig. 3b (evolution of the current with the nature of the counterions).



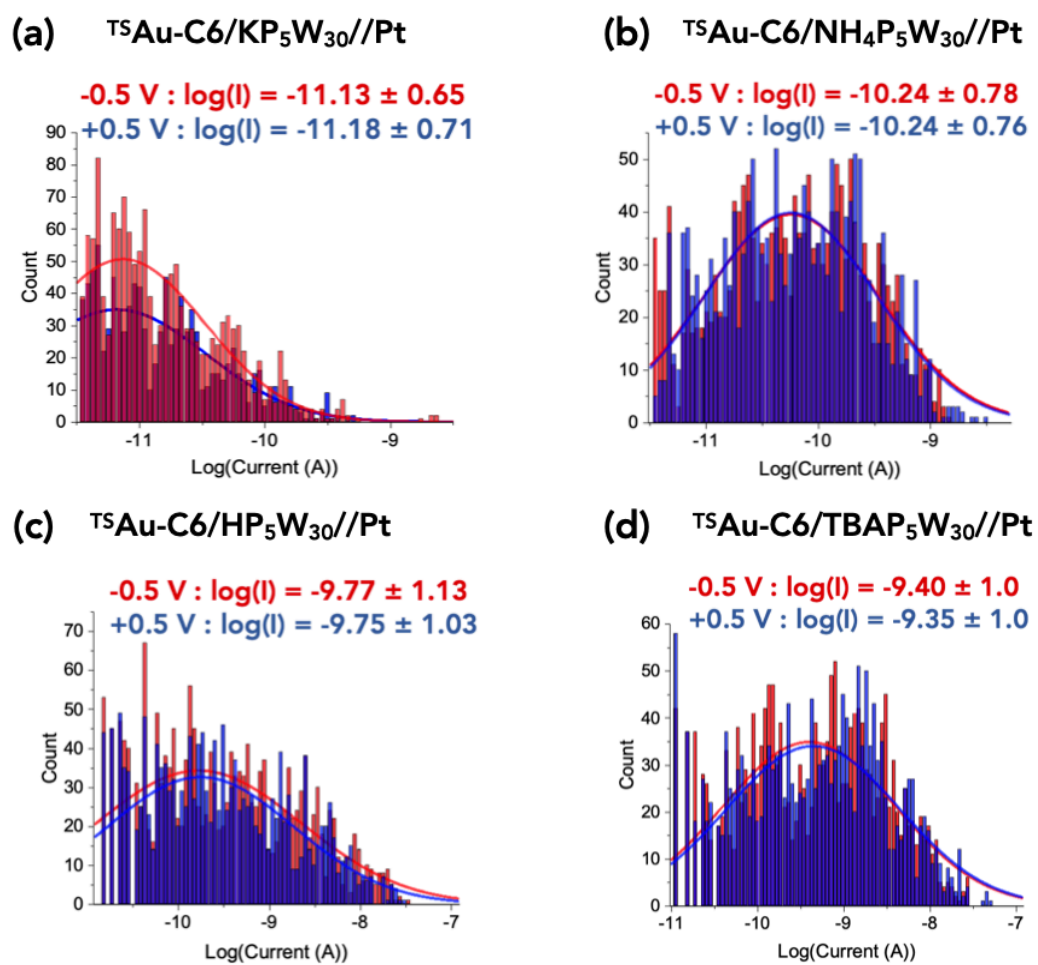

*Figure S9.* Histograms of log(current) measured at 0.5V and -0.5V for the four samples. The solid lines are the fit by a log-normal distribution with the values of the log-mean (log-$\bar{I}$) and a log standard deviation (log-$\sigma$) marked on the panels and summarize in Table S1.



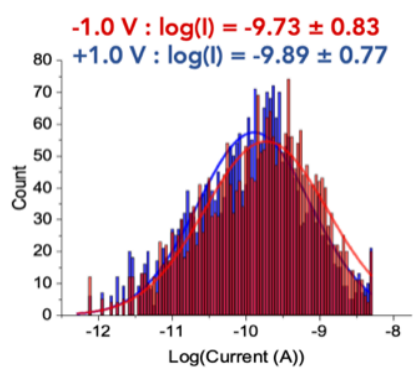
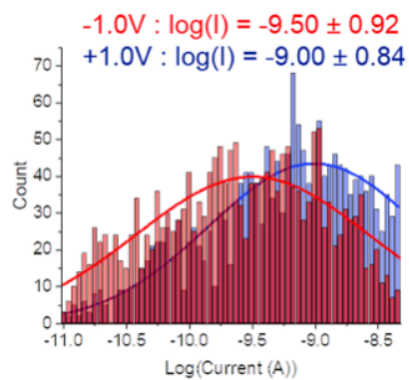
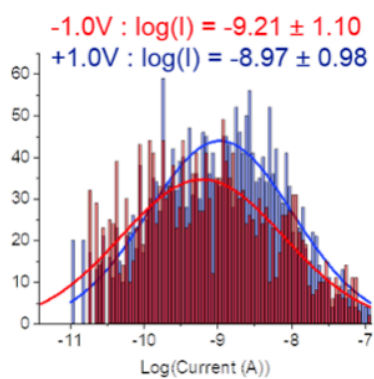
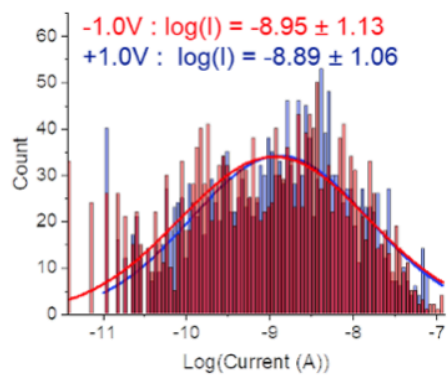

*Figure S10*. Histograms of log(current) measured at 1V and -1V for the four samples. The solid lines are the fit by a log-normal distribution with the values of the log-mean (log-Ī) and a log standard deviation (log-σ) marked on the panels and summarize in Table S1.



|  |  | KP$_5$W$_{30}$ | NH$_4$P$_5$W$_{30}$ | HP$_5$W$_{30}$ | TBAP$_5$W$_{30}$ |
|---|---|---|---|---|---|
| +0.5 V | log-Ī | -11.18 | -10.24 | -9.75 | -9.35 |
|  | Ī (A) | 6.6x10$^{-12}$ | 5.7x10$^{-11}$ | 1.8x10$^{-10}$ | 4.5x10$^{-10}$ |
|  | log-σ | 0.71 | 0.76 | 1.03 | 1.0 |
| -0.5 V | log-Ī | -11.13 | -10.24 | -9.77 | -9.4 |
|  | \|Ī\| (A) | 7.4x10$^{-12}$ | 5.7x10$^{-11}$ | 1.7x10$^{-10}$ | 4.0x10$^{-10}$ |
|  | log-σ | 0.65 | 0.78 | 1.13 | 1.0 |
| +1 V | log-Ī | -9.89 | -9.00 | -8.97 | -8.89 |
|  | Ī (A) | 1.3x10$^{-10}$ | 1.0x10$^{-9}$ | 1.1x10$^{-9}$ | 1.3x10$^{-9}$ |
|  | log-σ | 0.77 | 0.84 | 0.98 | 1.06 |
| -1 V | log-Ī | -9.73 | -9.50 | -9.21 | -8.95 |
|  | \|Ī\| (A) | 1.9x10$^{-10}$ | 3.2x10$^{-10}$ | 6.2x10$^{-10}$ | 1.1x10$^{-9}$ |
|  | log-σ | 0.83 | 0.92 | 1.10 | 1.13 |

***Table S1***. *Values of log-mean (log-Ī), mean current Ī, and a log standard deviation (log-σ) of the current dispersion (log-normal distribution) shown in Fig. S10.*

**6. Single Energy-Level Model**

The single-energy level (SEL) model (Eq. 1 main text), considers that: i) a single molecular orbital (MO) dominates the charge transport, ii) the voltage mainly drops at the molecule/electrode interface and iii) that the MO broadening is described by a Lorentzian or Breit-Wigner distribution.[16, 17] The simple energy scheme (Fig. S11) is described by $\varepsilon_0$ the energy of the MO involved in the transport (with respect to the Fermi energy of the electrodes), $\Gamma_1$ and $\Gamma_2$ the electronic coupling energy between the MO and the electron clouds in the two electrodes, e the elementary electron charge, h the Planck constant and N the number of molecules contributing to the ET in the molecular junction (assuming independent molecules conducting in parallel, *i.e.* no intermolecular interaction[18-20]). Albeit this number can be estimated using mechanical models of



the tip/SAM interface in some cases when the Young modulus of the SAM is reasonably known,[14, 15, 21-23] this is not the case here for the POM/alkyl SAM system for which the Young modulus has not been determined. Consequently, we use N=1 throughout this work. This means that the $\Gamma_1$ and $\Gamma_2$ values are "effective" coupling energies of the SAM with the electrodes and they are used only for a relative comparison of the POM SAMs measured with the same C-AFM conditions in the present work and they cannot be used for a direct comparison with other reported data (as for example from single molecule experiments). We also note that the exact value of N has no significant influence on the fitted parameter $\varepsilon_0$. The fits were done with the routine included in ORIGIN software,[24] using the method of least squares and the Levenberg Marquardt iteration algorithm.

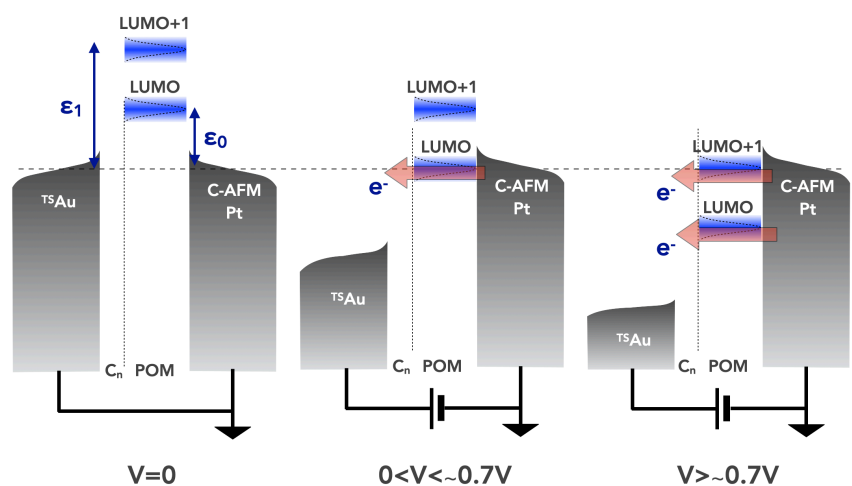

*Figure S11. Schematic energy diagram of the molecular junction at 0V, at a moderate positive voltage applied on the Au substrate V<0.6-0.7 V and at higher voltages V>0.7V.*



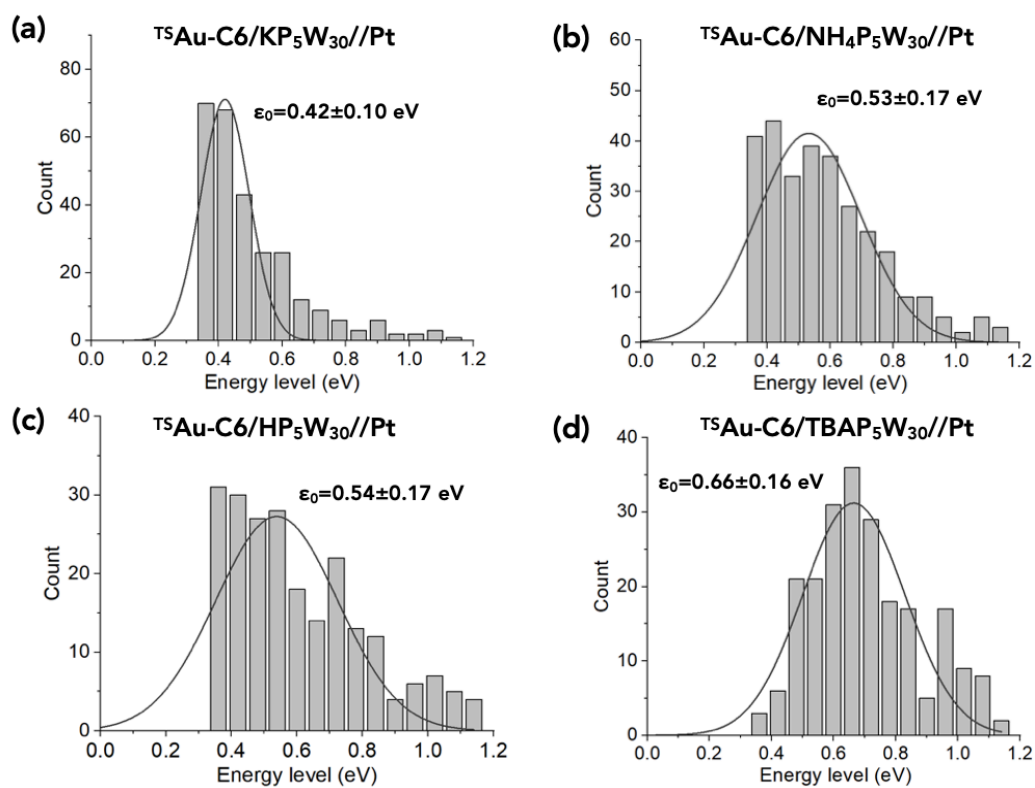

*Figure S12*. Histograms of the energy values $\varepsilon_0$ deduced from the fit of the SEL model on the complete I-V dataset shown in Fig. 2 (main text). The solid lines are the fits by a normal distribution, the mean values are shown in the panels.



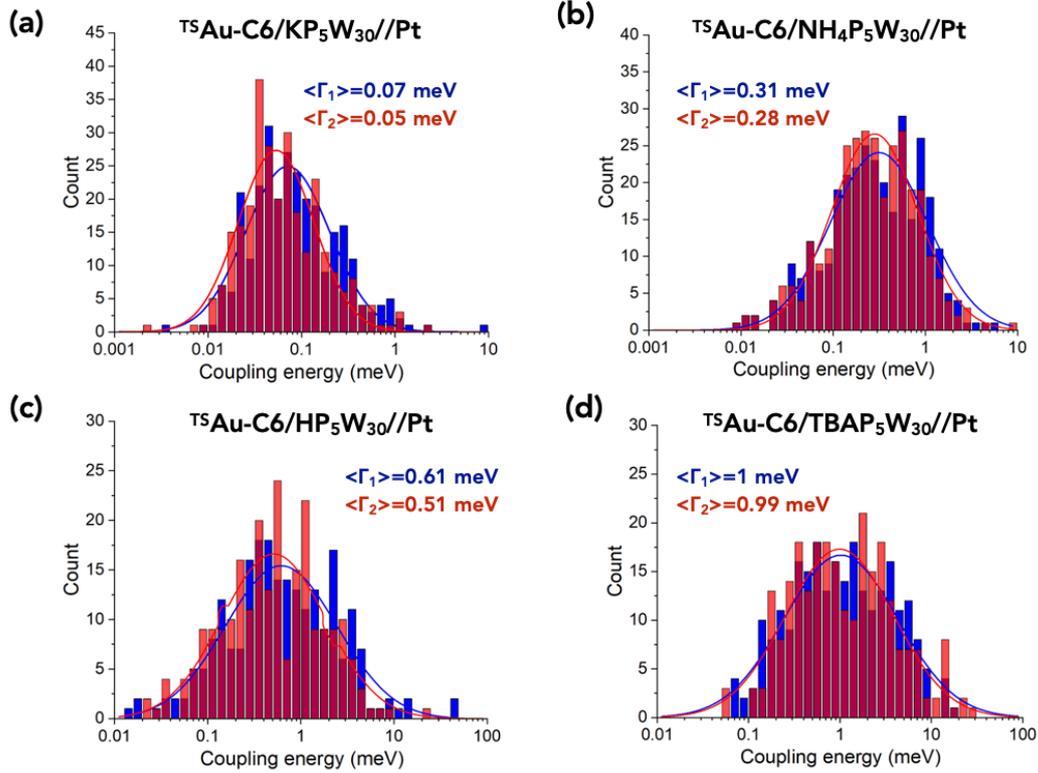

**Figure S13**. *Histograms of the coupling energy values $\Gamma_1$ and $\Gamma_2$ deduced from the fit of the SEL model on the complete I-V dataset shown in Fig. 2 (main text). The solid lines are the fits by a log-normal distribution, the mean values are shown in the panels.*

### 7. Transition Voltage Spectroscopy

We also used the transition voltage spectroscopy (TVS)[25-29] to analyze the I-V curves. Plotting $|V^2/I|$ vs. V (Fig. S14),[30] we determine the transition voltages $V_{T+}$ and $V_{T-}$ for both voltage polarities at which the bell-shaped curve is maximum. This threshold voltage indicates the transition between off resonant (below $V_T$) and resonant (above $V_T$) transport regime in the molecular junctions and can therefore be used to estimate the location of the energy level. In Fig. S14, the thresholds $V_{T+}$ and $V_{T-}$ are indicated by the vertical arrows (with values) and determined from the max of a 2nd order polynomial function fitted around the



max of the bell-shaped curves (to cope with noisy curves). The value of $\varepsilon_{0\text{-TVS}}$ is estimated by:[27]

$$\left|\varepsilon_{0-TVS}\right| = 2\frac{e\left|V_{T+}V_{T-}\right|}{\sqrt{V_{T+}^2 + 10\left|V_{T+}V_{T-}\right|/3 + V_{T-}^2}} \quad (2)$$

We also determined an average value of the electrode coupling energy $\Gamma_{TVS}$ using this relationship:[31, 32]

$$G(0) = NG_0 \frac{\Gamma_{TVS}^2}{\varepsilon_{0-TVS}^2} \quad (3)$$

with G(0) the zero-bias conductance (Fig. S15), $G_0$ the conductance quantum ($2e^2/h = 7.75 \times 10^{-5}$ S, e the electron charge, h the Planck constant) and N the number of molecules in the junction. G is calculated from the slope of the I-V curve in its ohmic region (-50 mV/50 mV) and N=1 (see above). Note that $\Gamma_{TVS}$ is equivalent to the geometrical average of the SEL values $(\Gamma_1\Gamma_2)^{1/2}$.[31, 32]

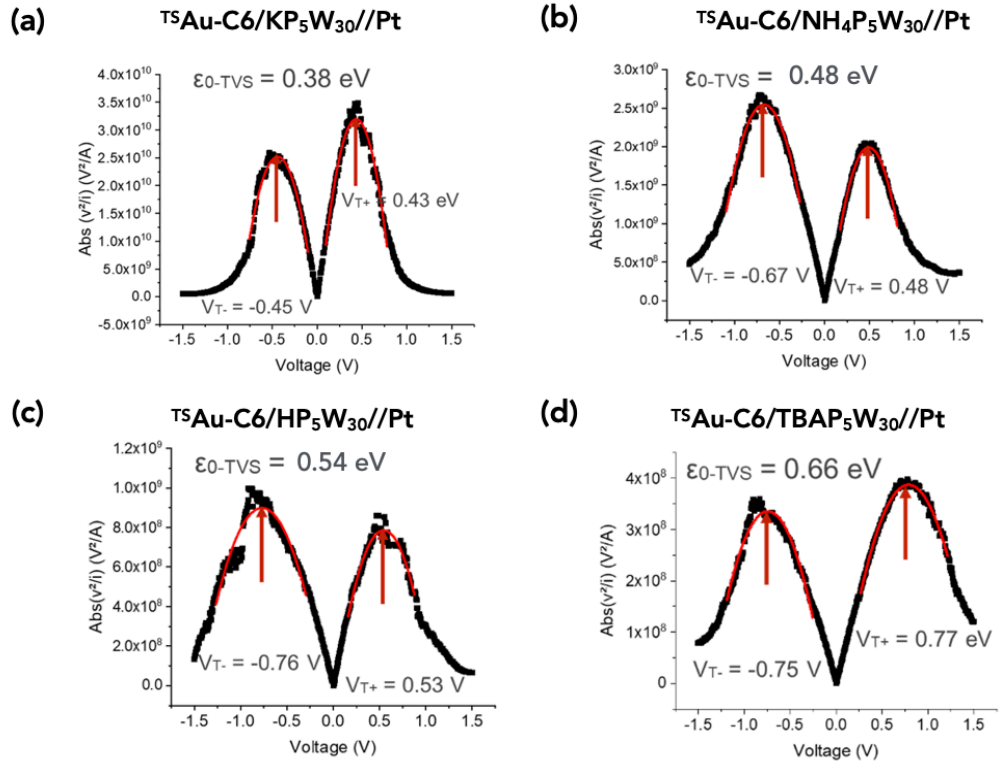

***Figure S14***. *TVS plot from the mean Ī-V traces (Fig. 2 in main text) for the four molecular junctions.*

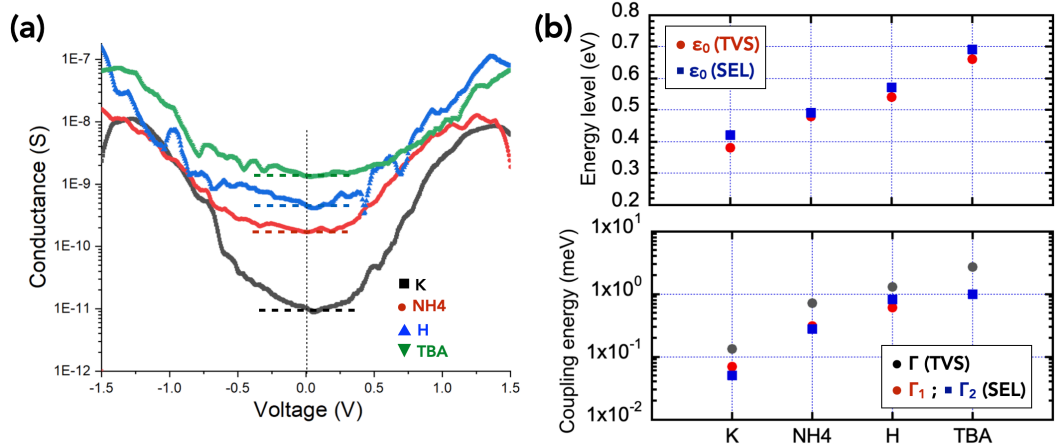

***Figure S15.*** *(a) First derivative (with Savitzky–Golay smooth on 20 data points) of the mean Ī-V to determine the zero-bias conductance (horizontal dashed lines)*

*and (b) comparison of the energy level and electrode coupling energies determined by the SEL model fit and the TVS method.*

**8. Supplementary figures for discussion**

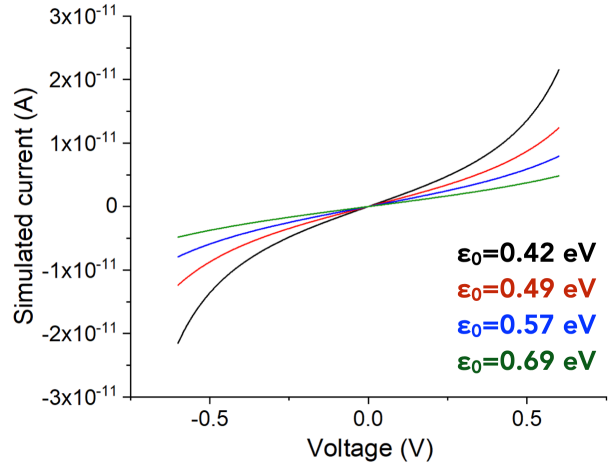

***Figure S16.*** *Simulated I-V (Eq. (1) in main text) with $\varepsilon_0$ ranging from 0.42 et 0.69 eV (see Table 2, main text) and keeping $\Gamma_1 = \Gamma_2 = 0.1$ meV, N=1. The ratio of the currents at +/- 0.6 V between the two extreme curves is 4.48.*

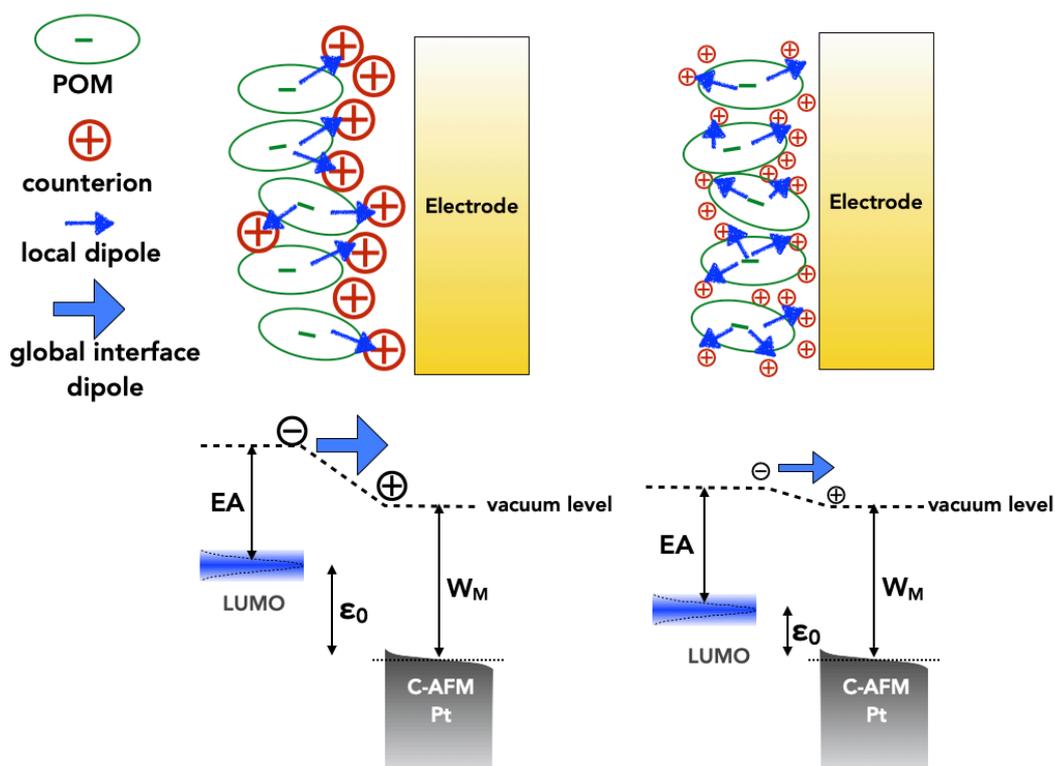

*Figure S17. Scheme of two possible scenarios at the POM/electrode interface for the TBA$^+$ counterions (left) and the smaller counterions, e.g. H$^+$, K$^+$ (right). The small blue arrows represent the local dipole between the POM and one counterion (not all represented for clarity). The large blue arrows indicate the resulting global interface dipoles that induce a vacuum level shift. For simplicity, we assume the same metal work function ($W_M$) and molecule electron affinity (EA) in the two cases. Due to the dipole-induced vacuum level shift, the LUMO is shifted upwards for the TBA$^+$ scenario (larger $\varepsilon_0$).*

**9. Two-sample t-test**

To determine if the small current asymmetry at +1V and -1V for the HP$_5$W$_{30}$ and NH$_4$P$_5$W$_{30}$ is statistically significant, we performed a two-sample t-test using the routine provided by ORIGIN software.[24] Since the t-test applies for normally distributed populations and the currents are log-normal distributed (Fig. S10), we consider the decimal logarithm of the currents taken from the datasets shown in Fig. 2 (main text) at -1V and +1V as the two populations 1 and 2 (P1, P2). We test the null hypothesis that the mean values are equal, mean(P1)-mean(P2)=0. At a significance level of 0.05, the data yielded a probability p-value that the null hypothesis is true of $4\times10^{-9}$ for $HP_5W_{30}$ and $6\times10^{-12}$ for $NH_4P_5W_{30}$, which are less than our 0.05 significance level, and thus we can reject the null hypothesis. In other words, the weak current asymmetry can be considered as statistically significant.